# Photon Diffusion in Microscale Solids

**Avijit Das, Andrew K. Brown, Merlin L. Mah and Joseph J. Talghader**

Department of Electrical and Computer Engineering, University of Minnesota-Twin Cities, 200 Union St SE, Minneapolis, MN 55455, USA

E-mail: joey@umn.edu

## Abstract

This paper presents a theoretical and experimental investigation of photon diffusion in highly absorbing microscale graphite. A Nd:YAG continuous wave (CW) laser is used to heat the graphite samples with thicknesses of 40 μm and 100 μm. Optical intensities of 10 kW/cm² and 20 kW/cm² are used in the laser heating. The graphite samples are heated to temperatures of thousands of kelvins within milliseconds, which are recorded by a 2-color, high speed pyrometer. To compare the observed temperatures, differential equation of heat conduction is solved across the samples with proper initial and boundary conditions. In addition to lattice vibrations, photon diffusion is incorporated in the analytical model of thermal conductivity for solving the heat equation. The numerical simulations showed close matching between experiment and theory only when including the photon diffusion equations and existing material properties data found in the previously published works with no fitting constants. The results indicate that the commonly-overlooked mechanism of photon diffusion dominates the heat transfer of many microscale structures near their evaporation temperatures. In addition, the treatment explains the discrepancies between thermal conductivity measurements and theory that were previously described in the scientific literature.

## 1. Introduction

The heat transfer and cooling of electronic and optical materials is generally treated as a combination of conduction, convection, and radiation; however, while these three mechanisms certainly dominate microscale systems near room temperature, they fail at extremely high temperatures. Under these conditions, near the evaporation temperature of many solids, photon diffusion dominates the heat transfer. In photon diffusion, thermal emission occurs throughout a material, but the photons produced are quickly reabsorbed and re-emitted repeatedly while traversing the material. This sequence of events creates a random walk of photons that can be modeled as a diffusion process. Photon diffusion has long been known to dominate the heat transfer in the interiors of stars [1, 2], but it has seldom been applied to solid materials, except in planetary interiors [3–5] and in diathermanous materials such as glasses, where heat transfer can occur via radiation through a transparent material but does not significantly heat the material itself [6–8]. This is unfortunate because there are many situations, such as particle combustion, laser machining, and even runaway device failure, where extremely high temperatures are reached, and traditional heat transfer models fail completely. It is noted that the term "photon diffusion" is often used to describe light scattering in biological tissues [9–11] and also studied to analyze the nonradiative energy transfer (i.e., Förster resonance energy transfer) in donor-acceptor fluorophore molecules [12–14]. However these applications do not imply photon diffusion as a heat transfer mechanism and therefore, are not further discussed in the paper.

In glass, radiant heat transfer was first reported by Kellett [6]. He mathematically explained the phenomenon of photon heat transfer inside diathermanous media like glasses. Later on, Genzel showed a quantification of photon heat conduction in terms of "photon thermal conductivity" for different glasses [7]. The photon conductivity explained the relation between radiative heat flux and temperature gradient and was subsequently used for other materials. In ceramics,

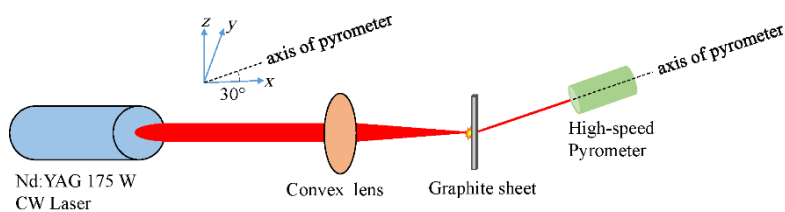

**Figure 1** Schematic illustration of the experimental setup. The collimated light coming from the Nd:YAG laser goes through the focusing lens and converges onto the graphite sheet producing a spot size of ~1mm. The light beam propagates along the $x$- direction and the graphite sheet is placed along the $y$-$z$ plane, normal to the light beam. A high speed, 2-color (ratio) pyrometer measures the transient temperature increase of the graphite sheet. For safety from laser damage, the pyrometer is placed at an angle of 30° inclined to the beam direction.

Lee and Kingery reported the theoretical and experimental measurements of photon conductivity for alumina, silica and Vycor-brand glass (V-1) [8]. They investigated the photon driven heat conduction from room temperature to 1500 K. In planetary rocks and minerals, photon conduction was extensively studied by Aronson et. al. [15] and Clauser [16]. Aronson et. al. calculated the photon conductivity for a number of minerals by measuring the refractive index and absorption coefficient at high temperatures. On the other hand, Clauser reported significant increase in the thermal conductivity (due to photon diffusion) of volcanic rocks at high temperatures (above 1473 K). In recent times, Keppler et. al. discussed the variation of the photon conductivity of Earth's lower mantle mineral (i.e., silicate perovskite) with pressures as high as 125 GPa [17]. However, all of the materials discussed above are either largely transparent or otherwise restricted to very large volumes. There is very little, if any, work reported on photon diffusion dominating thermal heat transfer on the microscale or in highly absorbing solids. Recently, the experiments of Mitra et. al. implied the dominance of photon diffusion for heat conduction in highly absorbing stainless steel particles under laser acceleration [18]. This work, however, was indirect, and encourages a direct measurement and theory of heat transfer in the high temperature microscale regime.

In this paper, we investigated the effect of photon diffusion to the heat transfer of highly absorbing microscale graphite. We used a continuous wave (CW) laser to heat up the graphite samples with intensities of 10 kW/cm$^2$ and 20 kW/cm$^2$. We measured the transient temperatures of the heated samples using a high speed pyrometer. In addition, we numerically solved the heat equation across the graphite samples considering both lattice and photon contributed heat conduction. Close matching was achieved between experiment and theory. Remarkably, the models only included

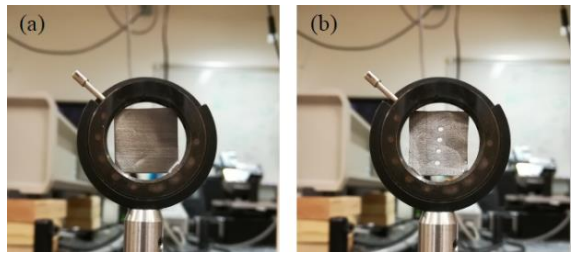

**Figure 2** (a) a 100 μm thick graphite sheet membrane suspended in a hollow sample aperture before laser illumination. (b) After 4 seconds of illumination, graphite sheet completely sublimates in the area of the incident laser beam. Four vertical holes, each of ~1 mm diameter, are created with 1–2 mm gaps between them. This clearly suggests that the sheet edge support does not affect the instantaneous heat transfer of the graphite sheet along its thickness.

fundamental photon diffusion theory and existing material properties for graphite found in published articles. No fitting constants were required to achieve matching.

## 2. Experimental setup

Laser heating experiments were performed at our main lab in the University of Minnesota, where a 175 W CW neodymium-doped yttrium aluminum garnet (Nd:YAG) laser was used to heat up the graphite samples. This laser was operated at a wavelength of 1064 nm. The output of the laser was fairly well collimated and was focused to produce a 1 mm beam spot. Output intensities of 10 kW/cm$^2$ and 20 kW/cm$^2$ were used for laser heating within the 1 mm spot size. The basic experimental setup is shown in Fig. 1. The collimated beam coming out of the Nd:YAG laser propagated along the $x$-direction and was focused through a convex lens onto a thermal graphite sample. The lens was placed 25 cm away from the laser and 16 cm away from the graphite sample. An exfoliated flexible graphite sheet from MinSeal was used as the graphite sample. The flexible sheet was manufactured from mineral graphite. Two sheet thicknesses were tested: 40 μm and 100 μm. The thicknesses of the sheets were quite uniform, with relatively small uncertainty of ~2%. Each sheet was cut into 2 cm × 2 cm square piece and the piece was suspended in the hollow aperture of a sample holder (i.e., optical mount) along the *y-z* plane, normal to the laser beam direction (Fig. 2). Under CW illumination, the transient temperatures of the heated graphite sheet were measured using a high speed pyrometer. The pyrometer was a 2-color, digital infrared device from Process Sensors and had a spectral response of 1.45 μm–1.8 μm. Since the pyrometer was not responsive to the operating wavelength (1.064 μm) of the Nd:YAG laser, no optical filter was used to block the Nd:YAG radiation. To protect from any damage incurred by the direct illumination of the Nd:YAG, the pyrometer was placed at an inclined angle of 30° respective to the *x*- direction. Extreme care was taken to focus the Nd:YAG beam spot and pyrometer alignment laser spot at the same point on two opposite faces of the graphite sheet. A slight misplacement of these spots would incur errors in the temperature measurements. After focusing, the pyrometer took transient temperature readings in the range of 1100 K–3000 K (measurement limit) at a response time of 3 ms. During the experiment, we found the graphite sheet to sublimate within ~3–4 seconds after the Nd:YAG illumination began. Fig. 2(a) shows a 100 μm thick graphite sheet (2 cm × 2 cm membrane) before laser illumination, and Fig. 2(b) shows the complete sublimation of the graphite sheet in the areas of the incident laser beam (1 mm beam spot) after 4 seconds of laser illumination. Therefore, to avoid significant material evaporation we limited all laser shots to ~1–2 seconds.

## 3. Theoretical development

### 3.1 Thermal model of heat transfer

In Fig. 3, we illustrate the concept of laser heating in a microscale graphite sheet. A graphite sheet with thickness L, is placed normal to the propagation direction of a Nd:YAG beam. The thickness of the sheet is along the *x*-axis, and the surface area is across the *y-z* plane. The laser beam illuminates the front surface (*x*=0) of the graphite sheet, producing an incident intensity $Q_{in}$ across a 1 mm beam spot. This initiates thermal heat conduction into the graphite sheet. However, the thickness of the sheet (40 μm or 100 μm) is so small compared to the surface area (2 cm × 2 cm) and the 1 mm beam spot that, we can ignore heat transfer across the surface (*y-z* plane) and consider one dimensional (1-D) heat penetration only along the thickness (*x*-axis), as seen in Fig. 3 [19]. In addition, we can ignore any heat conduction between the graphite and the sheet edge supports, as the incident beam is far away from those supports, and heat penetration along the thickness is unaffected by the sample holder (Fig. 2 (b)). After illumination, the heat flux is absorbed and then radiated from the back surface (*x*=L) of the graphite sheet, producing a radiant heat intensity $Q_{out}$. It is noted that we assume the scattering and hence, reflectivity from the front surface (*x*=0), to be negligible, and the absorption of incident heat flux to be 100% due to high power laser illumination [20, 21]. However, to find out the increase in temperature *T* at any time (*t*>0) or distance (0<*x*<L), the heat equation needs to be solved along the *x*- direction and can be expressed as [22, 23],

$$\rho C_P \frac{\partial T}{\partial t} = K \frac{\partial^2 T}{\partial x^2} \tag{1}$$

where $\rho$ is the density, $C_P$ is the specific heat at constant pressure, and *K* is thermal conductivity of graphite. To solve Eq. (1), proper boundary conditions must be applied on the front and back surfaces of the sheet, which requires modeling $Q_{in}$ and $Q_{out}$. $Q_{in}$ can be modeled using a one dimensional form of Fourier's Law as [24],

$$Q_{in} = -K \frac{dT}{dx} \tag{2}$$

where *dT/dx* is the temperature gradient along the thickness of the graphite sheet. On the other hand, $Q_{out}$ can be modeled using the Stefan-Boltzmann Law as [25, 26],

$$Q_{out} = \varepsilon \sigma T^4 - \varepsilon \sigma T_{en}^4 \tag{3}$$

where $\varepsilon$ is the emissivity of graphite, $\sigma$ is the Stefan-Boltzmann constant, and $T_{en}$ is the temperature of the surrounding environment. All these parameters are mentioned with their respective values in Table 1. It is noted that Eq. (3) is valid for a graphite sheet as graphite is a well-known blackbody material [27] and is commercially used in blackbody walls [28, 29] and infrared (IR) filaments [30, 31]. Moreover, micro/nanoscale carbon-based structures have been previously reported as blackbody absorbers [32].

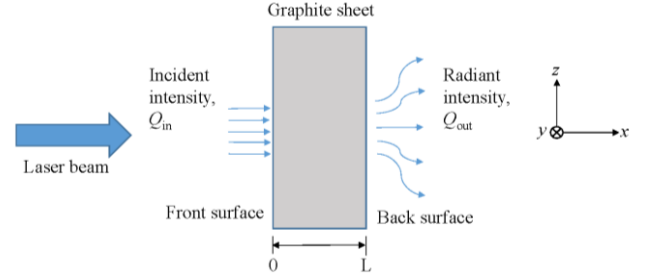

**Figure 3** Schematic illustration of the 1-D heat transfer in thermal graphite sheet. Nd:YAG laser beam produces an incident intensity, $Q_{in}$ on the front surface (*x*=0). The incident flux is absorbed and then radiated from the back surface (*x*=L), producing a radiant intensity, $Q_{out}$.

## 3.2 Thermal properties of graphite

Heat conduction in the microscale graphite sheet of course depends on the two major thermal properties of the material: thermal conductivity $K$ and specific heat $C_P$. Over the past few decades these properties have been analyzed and explained in the literature, along with their dependence on temperature [33, 34]. However, analytical expressions are required to describe the temperature dependency and hence, include in the thermal model which will be discussed in this section.

In graphite at very low temperatures (i.e., 50 K), heat conduction primarily depends on specific heat [35]. The thermal conductivity gradually increases with increasing temperature up to a maximum point. The temperature of this maximum point (also known as peak temperature) may vary from 80 K to 200 K, depending upon the crystalline structure of graphite [35, 36]. However, above the peak temperature and up to ~1800 K, heat conduction is dominated by lattice vibrations, or phonons, which show an inverse temperature dependence [35, 37, 38]. The phonon-phonon and phonon-defect scattering in a nonmetallic crystal, such as polycrystalline graphite can be understood from the temperature dependence of the Lorenz function, which is directly proportional to the thermal conductivity. The Lorenz function shows a gradual decrease with increasing temperature, which justifies the inverse temperature dependence of phonon dominant thermal conductivity of graphite [38]. A number of articles have verified this temperature dependence by reporting experimental data specifically for polycrystalline (i. e., Acheson) graphite in the range of 273 K to 1800 K [38, 39]. These experimental values of lattice or phonon dominant conductivity can be curve-plotted, and an empirical analytical expression can be formed as,

$$K_{lattice} = 230.114e^{-0.001204T} \qquad (4)$$

where temperature $T$ must be greater than or equal to 273 K. As room temperature is considered as the initial condition in our model, analytical modeling of $K_{lattice}$ below 273 K is not required. Note that exfoliated graphite is typically polycrystalline [40–42], and has similar physical and thermal properties as of Acheson graphite [43, 44]. Therefore, the experimental conductivity values of Acheson graphite can be used to model $K_{lattice}$. In addition, due to the empirical fitting of $K_{lattice}$, some errors might arise from the experimental measurements. From refs. [38, 39], these errors are found ±1–±4 Wm$^{-1}$K$^{-1}$ (normalized RMS error of ~3.8%) for temperatures up to 1800 K which are quite small, and results in the analytical expression of $K_{lattice}$ to be a good empirical model for lattice conductivity with adjusted R$^2$ of 98.2%. However, these small errors can be attributed to slight variations in the experimental measurements of the phonon conductivity, which originates from the random distribution of the crystal defects (i.e., grain boundaries, dislocations) in the polycrystalline graphite [45].

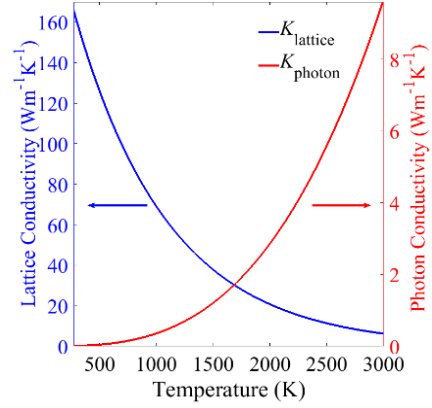

**Figure 4** Lattice and photon thermal conductivity of graphite with temperatures, as calculated from Eqs. (4) and (5), respectively.

At relatively high temperatures (above ~1800 K), photon diffusion is likely to contribute to the heat transfer through the sample in addition to lattice vibrations [46, 47]. The rate of diffusion (in other words, emission or absorption) of photons largely depends on the optical constants, i.e, refractive index and photon mean free path of graphite. The photon mean free path is comparatively smaller than the graphite sheet, therefore the path of photons resembles a random diffusion walk across the thickness of the sheet. In literature, the photon diffusion has been incorporated to the thermal conduction as "photon thermal conductivity". In general, the photon conductivity for graphite can be expressed as [1],

$$K_{photon} = \frac{16n^2\sigma T^3}{3\alpha} \qquad (5)$$

where $n$ is the refractive index and $\alpha$ is the mean absorption coefficient of graphite. These parameters may vary with temperature; however, the variations are found to be quite negligible (~0.43%–0.54% for $n$ and ~0.34%−1.2% for $\alpha$) [48]. Therefore, the temperature variations of $n$ and $\alpha$ are discarded in our thermal model.

Figure 4 presents a comparison between the lattice and photon thermal conductivities, as calculated from Eqs. (4) and (5) respectively. With increasing temperatures, $K_{lattice}$ decreases and $K_{photon}$ increases mainly due to their opposite temperature dependence. It is noted that lattice conductivities calculated from $K_{lattice}$ closely match with the ab-initio calculations derived for polycrystalline graphite (small grains) [49], with a normalized RMS error of ~4.5%. This further verifies the accuracy of the empirical modelling of $K_{lattice}$. An expression for $K_{photon}$ is already derived from first principle calculations [50, 51].

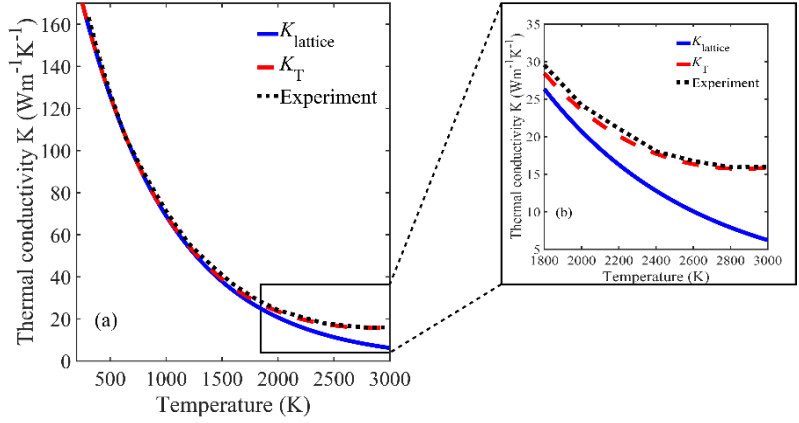

**Figure 5** (a) Thermal conductivity of graphite with temperatures in the range of 273 K to 3000 K. Eqs. (4) and (6) are simulated for $K_{lattice}$ and $K_T$, respectively. Experimental data are taken from [38]. (b) Thermal conductivities at high temperatures (1800 K–3000 K) are zoomed in on an inset to show the deviation between theory and experiment. It is noted that the deviations between $K_{lattice}$ and experiment were not covered in the previous articles and only our addition of photon diffusion contribution to the thermal conductivity ($K_T$) closely fits the experimental data.

At very high temperatures, the total thermal conductivity of graphite can be assumed as the summation of lattice and photon conductivity [8], that is

$$K_T = K_{lattice} + K_{photon} \qquad (6)$$

Note that electrons in polycrystalline graphite do not contribute significantly to the total thermal conductivity even at elevated temperatures. Using Wiedemann-Franz law [52] (which applies for a free electron gas such as the conduction electrons of graphite at high temperature) and experimental measurements of the electrical conductivity of polycrystalline Acheson graphite [38], the electron thermal conductivity is found to be ~5.58% of the lattice conductivity and ~5.55% of the photon diffusion conductivity even at elevated temperature (i.e., 2800 K). Therefore, the electron contribution to the thermal conductivity is neglected in Eq. (6).

At temperatures above 700 K, graphite is likely to oxidize in ambient air. Here the graphite reacts with oxygen and produces carbon monoxide (CO) and carbon dioxide ($CO_2$). This chemical reaction can release additional exothermic heat with an amount proportional to the oxidation reaction rate. In previous literature, the oxidation rate was found to be dependent on the air velocity when the temperature exceeds 1150 K [53], with a proportional increase in the oxidation rate at elevated temperatures (1150 K and above) when the air velocity is fixed [53]. Since the standard air velocity of the indoor environment (i.e., office, research space) lies between 15 cm/s to 22 cm/s [54, 55], we considered this range to calculate the oxidation rates at temperatures 1150 K and above. With the calculated rates, however, we analyzed the exothermic heat intensity produced on the graphite samples with formation of CO and $CO_2$ gases. Even at elevated temp-erature (i.e., 2800 K), the exothermic heat intensities were found only ~0.0115%–0.0119% of the 10 kW/cm² laser intensity when CO forms, and ~0.0433%–0.0448% of the 10 kW/cm² laser intensity when $CO_2$ forms. It is obvious that the exothermic heat intensities are quite negligible compared to the incident laser intensities (10 kW/cm² and 20 kW/cm²) used in the experiment. Therefore, the exothermic heat release, and hence the oxidation effects are not included in our thermal model. Moreover, the heat transfer between air and graphite is ignored, as it is quite negligible (~1.2%) compared to the heat transfer along the thickness of the graphite sheet even at high temperatures [56].

To further illustrate the validity of Eq. (6), we simulated thermal conductivity from Eqs. (4) and (6) for a wide range of temperatures (273 K–3000 K). Fig. 5(a) presents the simulated results along with the experimental data taken from [38], which had unexplained discrepancies between thermal conductivity models and data. It is noted that both theoretical and experimental conductivities show a gradual decrease with increasing temperature. At temperatures below 1800 K, $K_{lattice}$ is dominant and predicts the experimental results of the thermal conductivity quite well. However, at temperatures above 1800 K, experimental data start to deviate significantly from the analytical values of $K_{lattice}$. To further illustrate this scenario, thermal conductivities at high temperatures (1800 K–3000 K) are zoomed in Fig. 5(b). It is observed that with increasing temperatures, experimental data show a comparatively smaller decrease than $K_{lattice}$, resulting in an increasing relative error up to ~53%. However, we included $K_{photon}$ in the total thermal conductivity $K_T$ and found a close match with the experimental data points, providing much smaller relative error up to ~2.7%. We note that at relatively high temperatures, i.e., above 1800 K, $K_{photon}$ becomes significant and starts to dominate over $K_{lattice}$ mainly due to the large $T^3$ dependence of $K_{photon}$. This addition appears to fully explain the previous disagreement between prior theory and modeling for high temperature graphite.

**Table 1.** Parameters used in the numerical simulation

| Notation and meaning | Value of the parameter and unit |
| --- | --- |
| L, thickness of the graphite sheet | 40 μm and 100 μm |
| $Q_{in}$, optical intensity | 10 kW/cm$^2$ and 20 kW/cm$^2$ |
| $\sigma$, Stefan Boltzmann constant | 5.67x10$^{-8}$ W/m$^2$-K$^4$ |
| $\varepsilon$, emissivity of graphite | 0.9 [57] |
| $n$, refractive index of graphite | 3.45 [58] |
| $T_{en}$, temperature of surrounding environment | 300 K |
| $T_0$, room temperature (initial condition) | 300 K |
| $\alpha$, absorption coefficient | 100 cm$^{-1}$ [59] |
| $\rho$, density of graphite | 1.8 g/cm$^3$ |

In order to model the specific heat $C_P$, experimental measurements can be adapted from a number of research articles [60–65]. Due to the large number of types of graphite (i.e., natural, artificial, reactor grade, synthetic, etc.), not all of these measurements relate to the polycrystalline graphite. However, due to the nature of heat capacity and the reasonably similar densities of the different types of graphite, the specific heat is often assumed to be the same for all types [66]. Therefore, experimental evaluations for different ranges of temperatures have been combined, and a polynomial expression can be formed by curve-fitting the measurements as,

$$C_P = 2251.58 + 3.81 \times 10^{-2}T - 3.77 \times 10^5 T^{-1} - 1.82 \times 10^8 T^{-2}$$
$$+ 6.66 \times 10^{10}T^{-3} - 6 \times 10^{12}T^{-4} \qquad (7)$$

where temperature $T$ lies in the range of 200 K to 3000 K. It is noted that Eq. (7) is an empirical fit to experimental data measured by other research groups; therefore, there are some errors associated from the experimental measurements of the specific heat. From ref. [66], the associated errors are found as ±15–±27 Jkg$^{-1}$K$^{-1}$ for temperatures below 1000K, and ±29–±53 Jkg$^{-1}$K$^{-1}$ for temperatures above 1000 K. However, these errors are quite small compared to the analytical values of $C_p$ which makes Eq. (7) a good empirical fit for specific heat, with adjusted R$^2$ of 95.3% and 90.8% for temperatures below 1000 K and above 1000 K, respectively.

## 4. Results and discussion

To understand the heat transfer across the microscale graphite sheet, we solved the partial differential Eq. (1) with boundary conditions stated in Eqs. (2) and (3). Analytical expressions of $K$ and $C_P$ were used from Eqs. (4) – (7). For numerical simulation, we used the partial differential equation (PDE) toolbox in MATLAB. The PDE toolbox used numerical differentiation formulas (NDFs) of orders 1 to 5 to solve the heat equation [67]. Relative error tolerance was kept at 10$^{-5}$; therefore, the local discretization error was below this tolerance. We simulated the transient temperatures under the laser intensities of 10 kW/cm$^2$ and 20 kW/cm$^2$. To illustrate the effect of photon diffusion, we added $K_{lattice}$ and $K_{photon}$ to form the effective thermal conductivity $K_T$. We used both $K_{lattice}$ and $K_T$ into our simulation, and compared the simulated results with our experimental measurements. To ensure accuracy and precision of our experimental data, we measured three sets of transient temperatures for each experiment and then took the average to estimate the actual set of temperature measurements. The relative uncertainty of temperature measurements was found ~0.2%–0.5% which was quite negligible. As all the experimental measurements were taken from the back surface of the graphite sheet, the numerical simulations were also performed at $x$=L, as seen in Fig. (3).

Figure (6) shows the comparison of experimental and simulation results for 40 μm and 100 μm graphite sheets with laser intensity of 10 kW/cm$^2$. With the same illumination duration of 1000 ms, temperatures (theoretical and experimen-tal) in the 40 µm sheet show slightly larger increases than temperatures in 100 µm sheet. In addition, temperatures simulated with $K_T$ surpass those simulated with $K_{lattice}$ after reaching ~1800 K, mainly due to the increasing effect of $K_{photon}$. Therefore, conduction by photons produces a distinct temperature difference between the simulated results at 1000 ms. From Figs. 6(a) and 6(b), the difference is found as ~85 K and ~40 K for 40 µm and 100 µm sheets, respectively. However, in both cases, temperatures simulated with $K_T$ can consistently predict the experimental measurements, implying the contribution of photon diffusion to the heat conduction.

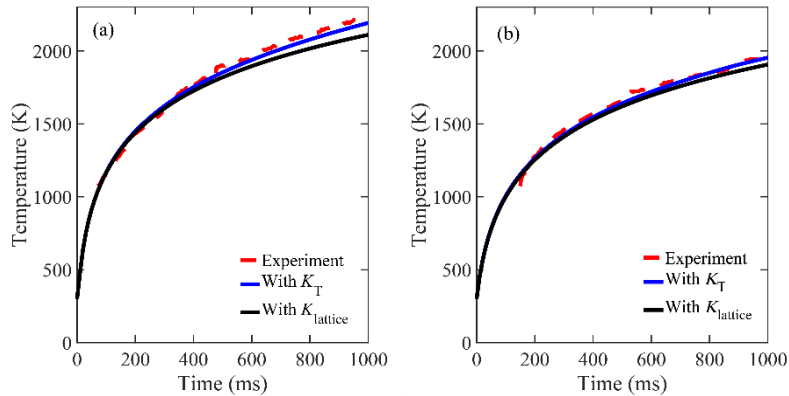

**Figure 6** Transient temperature profile for (a) 40 μm and (b) 100 μm thick graphite sheets with optical intensity of 10 kW/cm$^2$.

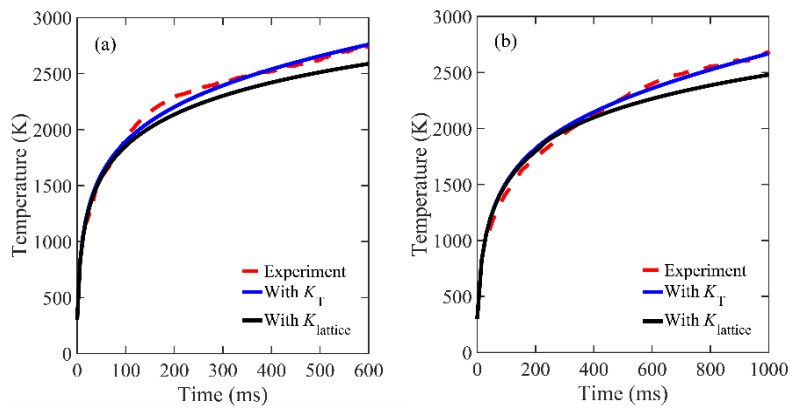

**Figure 7** Transient temperature profile for (a) 40 µm and (b) 100 µm thick graphite sheets with optical intensity of 20 kW/cm².

To further investigate the effect of photon diffusion, the laser intensity was increased to 20 kW/cm². Fig. (7) shows the simulated and experimented transient temperatures for both 40 µm and 100 µm graphite sheets with the intensity of 20 kW/cm². We note that during the experiments, temperatures in the 40 µm sheet increased much faster than those in the 100 µm sheet and surpassed the measurement limit of our pyrometer (3000 K) within 800 ms. Therefore, instead of taking same duration of 1000 ms for both thicknesses, we used different illumination durations. In Figs. 7(a) and 7(b), we show time durations of 600 ms and 1000 ms for 40 µm and 100 µm sheets, respectively. Due to the increased laser intensity in this set of experiments, the effects of photon diffusion can be seen more rapidly than in Fig. (6). Therefore, the difference between the simulated temperatures with $K_T$ and $K_{lattice}$ becomes larger than before. For the 40 µm sheet, the simulation difference is observed as ~170 K at 600 ms, whereas for the 100 µm sheet, the difference is found to be ~179 K at 1000 ms. We note that for 40 µm sheet, the difference would likely have been much larger than 170 K if measurements could have been performed at 1000 ms. However, the difference already proves the significance of photon diffusion at larger intensity. To further validate this idea, experimental measurements also show reasonable agreement with the temperature simulations performed with $K_T$, as shown in Fig. (7).

## 5. Conclusion

A detailed theoretical and experimental analysis was performed to study the effects of photon diffusion on the heat transfer in microscale graphite sheets. A Nd:YAG CW laser with laser intensities of 10 kW/cm² and 20 kW/cm² was used to heat 40 µm and 100 µm thick graphite sheets. Temperatures were found to rise up extremely high (2000 K and above) within milliseconds, and these temperatures were recorded by a 2-color, high speed pyrometer. To compare the experimental measurements, a thermal model of heat conduction was developed across the thickness of the graphite sheet. Necessary thermal properties, thermal conductivity and specific heat were analytically modeled to solve the heat equation across the sheet. When both photon diffusion and lattice vibrations were incorporated in the thermal conductivity of graphite, the results matched well with previous experimental data. However, to have a better understanding, the transient temperatures were numerically solved in two ways; one with only lattice conductivity and the other with combined lattice and photon conductivity. Significant temperature differences were observed between the two simulations with time. However, temperatures simulated with combined conductivity were found in good agreement with our experimental measurements, as expected from our earlier analysis of thermal conductivity. This validates the contribution of photon diffusion to the heat conduction in high absorbing microscale graphite.

## Acknowledgements

The authors would like to thank the Directed Energy Joint Transition Office (DEJTO) and the Office of Naval Research for support under grants N00014-17-1-2438 and N00014-12-1-1030.